\documentclass{jetpl}

\twocolumn

\lat

\DeclareMathOperator{\Tr}{Tr} 

\DeclareMathOperator{\al}{\alpha}
\DeclareMathOperator{\ep}{\epsilon} 
\DeclareMathOperator{\si}{\sigma}
\DeclareMathOperator{\vep}{\varepsilon} 
\DeclareMathOperator{\vk}{\varkappa} 
\DeclareMathOperator{\lm}{\lambda} 
\DeclareMathOperator{\om}{\omega} 
\DeclareMathOperator{\h}{\hbar }  
\DeclareMathOperator{\vd}{\varDelta}  

\DeclareMathOperator{\lala}{\langle \! \langle} 
\DeclareMathOperator{\rara}{\rangle \! \rangle} 

\DeclareMathOperator{\pa}{\partial} 
\DeclareMathOperator{\pr}{^{\prime}} 
\DeclareMathOperator{\ppr}{^{\prime \!\prime}} 
\DeclareMathOperator{\gx}{\nabla_{\! X}} 
\DeclareMathOperator{\gy}{\nabla_{\! Y}} 
\DeclareMathOperator{\gv}{\nabla_{\! V}} 
\DeclareMathOperator{\gp}{\nabla_{\!\! p}} 

\DeclareMathOperator{\ol}{\widehat{L}}
\DeclareMathOperator{\os}{\widehat{S}}
\DeclareMathOperator{\ok}{\widehat{K}}
\DeclareMathOperator{\oev}{\widehat{E}} 
\DeclareMathOperator{\ov}{\widehat{V}} 
\DeclareMathOperator{\oy}{\widehat{Y}} 

\DeclareMathOperator{\mcl}{\widehat{\mathcal{L}}}
\DeclareMathOperator{\mcs}{\widehat{\mathcal{S}}}
\DeclareMathOperator{\mck}{\widehat{\mathcal{K}}}

\DeclareMathOperator{\mct}{\widehat{\mathcal{T}}} 
\DeclareMathOperator{\mci}{\widehat{\mathcal{I}}}
\DeclareMathOperator{\mcu}{\widehat{\mathcal{U}}}

\title{On 1/f-noise of electron in phonon field}

\rtitle{On 1/f-noise of electron in phonon field}

\sodtitle{On 1/f-noise of electron in phonon field}
 
\author{Yu.\,E.\,Kuzovlev\/\thanks{kuzovlev@kinetic.ac.donetsk.ua, 
yuk-137@yandex.ru}}

\rauthor{Yu.\,E.\,Kuzovlev}

\sodauthor{Yu.\,E.\,Kuzovlev}

\address{Donetsk Free Statistical Physics Laboratory}

\dates{3 April 2017}{*}

\abstract{Exact propagator of density matrix of particle (electron) 
under influence of thermal vibrations of its medium (phonons) 
is treated in simplest approximation beyond the Fermi's golden rule. 
It is shown that uncertainties \,$\sim\h/t$\, in energy exchanges with 
the medium give rise to flicker fluctuations in rate of diffusion 
(diffusivity) of the particle.}


\begin{document}

\maketitle

{\bf 1}. One of actual themes of statistical mechanics is effects 
lost under its standard kinetic coarse-graining, 
and first of all thermal 1/f noise  (see \cite{ufn}-\cite{gs}) 
thus turned into a mystery like ``new phlogiston''. 
In particular, quantum kinetics essentially exploits the 
``Fermi's golden rule'' \cite{vk} to break continuous interactions 
of particles and quanta to mutually decoherent events.  
Corresponding theory's deficiencies by their nature are insensible 
to intensity/weakness  and details of interactions and therefore 
may be learned by examples of model mechanical systems  
of many particle (degrees of freedom).

{\bf 2}. In this respect it is  principally interesting to consider   
 {\sl \,a particle (``electron'') in potential field of vibrations 
 of surrounding medium (``phonons'')\,} with simple Hamiltonian  
\,$H=H_e +H_{ph}+H_{int}$\, where 
\begin{align}  \label{h}
H_e =\frac {p^2}{2m} \,\,, \,\,\,\, H_{ph}= \oint   
\hbar \omega_k\, a_k^\dag a_k\,\,,  \nonumber \\
H_{int}=  \oint  c_k\,  
( e^{\,ikr}a_k +  e^{-ikr} a_k^\dagger) \,\,, 
\end{align} 
\,$\oint \dots\,\equiv\int \dots\, d^3k/(2\pi)^3$\,, 
\,$c_k=c_k^*$\,, \,$|c_k|=c(|k|)$\, and  
\,$\om_k \hm = \om (|k|) >0$\,. 
There various phonon modes (PM) do not interact one with another, 
but this is unimportant for the electron (e.) since variety of 
modes is innumerably infinite and hardly e. could exchange 
momentum-energy with some of them more than once. 
Therefore if PM occupation statistics initially (let at \,$t=0$) 
is equilibrium then from e.'s viewpoint it henceforth will be the same. 

We are interested in statistics of fluctuations and relaxation of  
of e.'s velocity \,$V=p/m$\, and its ``Brownian walk'' in the space.  
 In  \cite{oct} it was already pointed out that the walk 
 statistics must be  radically non-Gaussian by containing    
 ``flicker'' fluctuations of e.'s mobility, since their absence would 
 be incompatible with unitarity of evolution of the system (\ref{h}) 
that is with that spectrum of its Liouville (von Neumann) 
operator is purely imaginary. 
Now we want to go from this ``existence theorem'' and exact equations 
for full system's density matrix (DM) \,$\rho_{e-ph}(t)$\,
and marginal e.'s DM \,$\rho (t) \hm =\Tr_{ph}\,\rho_{e-ph}(t)$\, 
to an approximation being as simple as possible 
but able to catch the flicker noise.

 {\bf 3}.\, With this purpose it is comfortable to describe e. in the Wigner 
 representation in terms of 
 \[
 \rho (t,X,Y)= \langle X+Y/2 |\,\rho (t)\,| X-Y/2\rangle \,\,,  
 \] 
 phonons in the coherent states representation, and them together
 via characteristic functional of electron-phonon (e.-ph.) correlations  
\begin{eqnarray}
 \label{gf}  
 \Tr_{ph}\, \{ \exp{[\int_k  z_-(k) \,a_k e^{ikr\pr} ]}\, 
  \,\times \\ \times\,\, \nonumber \, \langle r\pr | \,\rho_{e-ph}(t)\, |r\ppr\rangle \,  
\exp{[\int_k z_+(k) \,a_k^\dag e^{-ikr\ppr} ]} \} \,\, , 
\end{eqnarray}
where \,$r\pr =X+Y/2,\,r\ppr = X-Y/2$\, and \,$\int_k\dots\,=\int \dots\, d^3k$\,. 
With its means in \cite{july,oct} (our ``supplement materials'')  
electron propagator (EP) was obtained and expressed by
\begin{eqnarray} 
\rho (t)  =  \os(t)\,\rho (0)  \equiv  \,\,\,\,\,\,  \label{ep} \\    \nonumber \,\,\,\, \equiv  
 [\, \exp{\{\,t\,(- \ov \gx  + \mcl(z) )\}}\, \rho (0) \, ] |_{z =0}  \, \,   
\end{eqnarray} 
with e.'s velocity operator \,$\ov \hm = - i\,(\h/ m) \gy$\, and  
the Liouville operator in the form 
\begin{eqnarray} 
\mcl(z)  = i\sum_{+,-} \pm \int_k 
[\,(\om_k - k\ov) \,z_\pm (k)\, \frac {\delta}{\delta z_\pm(k) }  - \, \,\,\, \label{l} \\ - 
\,\frac {c_k}\h  \, B_{k} (Y)\, z_\pm(k)\, 
  + \frac  {c_k} {(2\pi)^3\h} \, A_{k}(Y)\,\frac {\delta}{\delta z_\pm(k) } \,] \,\,  \nonumber
\end{eqnarray}
with correlations birth-annihilation amplitudes
\begin{eqnarray}
B_{k} (Y)\equiv e^{\,ikY/2}N_k - e^{-ikY/2}(N_k+1) \, \, , \label{ab}  \\ \nonumber    
A_{k} (Y)\equiv e^{\,ikY/2}-e^{-ikY/2} \,\, ,   
\end{eqnarray} 
where \,$N_k = [\exp{(\h \omega_k/T)} -1]^{-1}$\,.

{\bf 4}.\, Let us write out EP \,$\os(t)$\, evidently.  
Since operator \,$\mcl(z)$\, is of first-order in respect to differentiations 
\,$\delta/\delta z_\pm (k)$\, the expression (\ref{ep}) would 
be easy calculable if this was not prevented by  
non-commutativity of \,$\ov$\, and \,$Y$\, in \,$\mcl(z)$\,. 
We can avoid this difficulty by assigning to  \,$\ov$\, and \,$Y$\, 
auxiliary time arguments and making the exponential 
chronologically ordered by it. 
Keeping an eye on it gives possibility to forget the non-commutativity 
for a time and easily transform the exponential to the normal 
form in respect to \,$\delta/\delta z_\pm (k)$\, and \,$z_\pm (k)$\,. 
Thus one finds 
 \begin{eqnarray} 
 \label{epe}   \os(t)\,= 
  \mct\,  \exp\,\{ \int_0^t [-\ov (t\pr)\gx  ]\,dt\pr  +  \, 
     \\  +    \nonumber
  \int_0^t \!\!\int_0^{t_1} \ok [t_1,t_2,Y(\cdot),\ov(\cdot) ] \,dt_2 dt_1\,\} \,\, ,  
\end{eqnarray}
where  \,$\mct$\, symbolizes chronological ordering and  
\begin{eqnarray} 
 \ok [t_1,t_2,Y(\cdot),\ov(\cdot) ]    =    \oint \,  \frac { |c_k|^2}{\h ^2}  \, A_{k}(Y(t_1))   \,   
  \times    \,\,\, \label{epker}   \\   \, \times  \, \nonumber     
 \,  2\, \cos\, \{  \int_{t_2}^{t_1} [\, \om_k - k\ov(t\pr) ]\,dt\pr\} \,  
 B_{k}(Y(t_2))  \,\,  .
 \end{eqnarray} 
 Of course, after \,$\mct$\,'s action the auxiliary arguments are removed. 
 
 {\bf 5}.\, Ignoring  \,$\mct$\, we come to approximation  
 \begin{eqnarray} 
 \os(t)  \Rightarrow   \os_{1/t}(t)  \equiv   \label{ap}  
  \exp{ \{ \,t\,[-\ov \gx +  \mck_{1/t}\, ] \} }  \,\, , \, 
 \\     \label{ko}  
  \mck_{1/t}  \equiv \frac 1t 
 \int_0^t \!\!\int_0^{t_1} \ok [t_1,t_2,Y,\ov] \,dt_2 dt_1   =  \, \\ \nonumber   = 
  \oint \,  \frac { |c_k|^2}{\h ^2}      
 \,  A_{k}(Y)    \,  \frac { 2\,[1-\cos\,[t\,(\om_k - k\ov) ] ]} 
  {(\omega_k - k\ov)^2 t} \,  B_{k}(Y)  \,\, .    
 \end{eqnarray} 
 Hear \,$\mck_{1/t}$\, appears as fictitiously time-local 
 kinetic operator (KO). In fact it is   
 ``pseudo-kinetic''  (PKO) since it depends on the whole 
 observation time. Introducing  \,$\nu=1/t$\, and 
 fuzzy ``quasi-delta-function'' (QDF)  
 \begin{align}  \label{kdf}
   \delta_\nu(\om) =   \frac { 2\nu\,\sin^2 (\om/2\nu)}  {\pi \om^2} \approx 
   \frac 1\pi\, \frac { \nu}  {\nu^2+\om^2} \,\,  
\end{align}
and going from  \,$\ov$\, and \,$Y$\, to Wigner-conjugated pair  
  \,$V$\, and \,$\oy \hm = i\,(\h/m)\gv $\, let us write shortly 
\begin{eqnarray} 
 \label{ko_}  
 \mck_{\nu}  = \oint \frac { 2\pi |c_k|^2}{\h ^2}\,
 \,  A_{k}(\oy)    \,  \delta_{\nu} (\om_k - kV) \, B_{k}(\oy)  \,\, .   
\end{eqnarray} 
Unfolding this expression, one can make sure that 
\begin{eqnarray}  \label{kok} 
\mck_{\nu} \rho(V) = 
\int_{V^\prime}  \,[w_{\nu}(V|V^\prime)\,\rho(V^\prime) 
- w_{\nu}(V^\prime|V)\,\rho(V)\,] \,\, 
\end{eqnarray}
with \,$\int_V \dots =\int \dots\, d^3V$\, and   
\begin{eqnarray}
w_{\nu}(V|V^\prime) = \frac {m^3}{ (2\pi)^2 \h^4} \,\,  \label{tp}
|c_{m(V -V\pr)/\h} |^2 \,\, 
\times \,\,\,\,\,  \\ \times  \,\,  \nonumber 
\{ \, N_{m(V -V^\prime)/\h}   \,\, 
\delta_{\hbar\nu}(\h\om_{m(V -V^\prime)/\h} - E + E^\prime) \, 
+ \, \\  \nonumber + \, 
[ N_{m(V^\prime-V)/\h} +1 ] \, 
\delta_{\hbar\nu}(\h\om_{m(V^\prime -V)/\h} + E - E^\prime) \}\,\,  
\end{eqnarray}
with  \,$E\equiv mV^2/2$\, and \,$E\pr\equiv mV^{\prime 2}/2$\,. 
According to (\ref{kok}) our PKO looks like 
Kolmogorov-Chapman type operators in formalism (``master equations'' \cite{vk})  
for Marcov stochastic processes. However, our transition 
rates  (``probabilities'') (\ref{tp}) are time dependent through 
the QDF's width \,$\nu=1/t$\, in such way that EP (\ref{ap}) is 
not representable as solution to a first-order time differential 
equation, which makes the theory strongly non-Marcovian.

{\bf 6}.\, If applying the ``golden rule'', that is replacing 
QDF in (\ref{ko_})-(\ref{tp}) by the usual Dirac's delta-function 
(DDF) \,$\delta (\om)$\, and thus turning PKO into KO,  
\begin{align} 
 \os(t)  \Rightarrow  \os_0(t)  \equiv  \exp{ \{ \,t\,[- V \gx +  \mck_0 ] \} }  \,\,,  \,  
 \,\,\,   \nonumber  \\    
 \mck_0 \equiv   \oint \frac { 2\pi |c_k|^2}{\h ^2}\,  A_{k}(\oy)  
 \, \delta(\om_k - kV)\, B_k(\oy) \,\, ,
 \label{uko}  
\end{align}
one comes to to the conventional (Marcovian) kinetics' approximation. 
KO (\ref{uko}) is \cite{july}  mere 
 ``one-electron version'' of the e.-ph. collision integral \cite{lp}.  
 It produces the Maxwell distribution as equilibrium one:   
 \begin{align}
 \mck_0 \, \rho_0(V) =0\,\,, \,\,\,\, 
 \rho_0(V) \propto \exp{(-mV^2/2T)} \,\, .    \label{eq}  
 \end{align}
It follows from easy justifiable equalities 
\begin{align}   
\delta(\om_k-kV)\, B_k(\oy)\,\rho_0(V)  = 0\,\, . \,  \label{db0} 
\end{align} 
Their validity for any PM separately says about  
``detail balance'' of  e.-ph. collisions. 

   
One may believe that at suitable parameters of (\ref{h})   
(may be with several PM branches \cite{july}) \,$\mck_0$\,       
ensures fast relaxation of e.'s velocity distributions and correlators. 
But this is certainly impossible in the more ``high'' approximation (\ref{ap})-(\ref{tp}).

{\bf 7}.\, Indeed, after returning QDF to its place instead of DDF 
the equalities (\ref{db0}) no more are valid, i.e. detail balance 
(DB) is not completely observed in finite time frameworks. 
Therefore a distribution \,$\rho_\nu(V)$\, stationary in respect 
to PKO (\ref{ko_})-(\ref{tp}) and satisfying    
 \begin{align}   \label{0}  
  \mck_\nu \rho_{\nu}(V)  = 0     \,\,\,\,\,\, \,\, (\nu = 1/t)\,\, ,\, 
\end{align} 
does not coincide with the Maxwellian \,$\rho_0(V)$\,. 
Their difference is determined by  ``wings'' of QDF.  
From (\ref{ko_}) we have   
\begin{eqnarray} 
\mck_{\nu} \rho_0 \approx    \nonumber 
- \,\nu   \oint \frac { 4 |c_k|^2}{\h ^2}\, N_k\, 
\exp\, [\, \frac {\h\om_k}{2T} - \frac {(\h k)^2}{8mT} \,]\,
 \times \\\times\,  \label{kor0}  
A_{k}(\oy) \,
\frac {\sinh\,[\h\, (\om_k - kV)/2T] } 
{ (\om_k - kV)^2} \, \rho_0(V)  \,\,   
\end{eqnarray} 
with integral in the principal value sense. 
Then from (\ref{0}) in the form      
\,$\mck_\nu [\rho_{\nu} -\rho_0 ] \hm + \mck_{\nu}  \rho_0 \hm = 0 $\,  
we see that difference \,$\rho_\nu-\rho_0$\, 
decreases with time not faster 
than  
\begin{align}    \label{1}  
  \rho_{1/t}(V)  -\rho_0(V) \, \propto \, 
  \frac {\tau(V)}t \,\rho_0(V)   \,\, \,    
\end{align} 
with some function \,$\tau(V)$\,. 
 
It shows that relaxation of arbitrary   
initial velocity distribution \,$\rho_{in}(V)$\, 
to equilibrium \,$\rho_{0}(V)$\, may go by  
very slow - time non-integrable - law. 
However,  origin of such behavior is not conservation 
of some physical quantity but DB's violation. 
Hence a slow scenario is not obligatory and there must 
exist initial conditions \,$\rho_{in}(V)$\, promoting DB and 
leading to equilibrium in a fast (integrable) way. 
In such case in identity 
\begin{align}   \nonumber   
\begin{array}{c}
\os_{1/t}(t) \, \rho_{in}(V) - \rho_0(V)  = 
\os_{1/t}(t) \,[ \rho_{in}(V) - \rho_0(V) ]\,  + \, 
\, \\ + \, 
[\,1 -\os_{1/t}(t) ] \,[\rho_{1/t}(V) -\rho_0(V) ] \,  
\end{array}
\end{align} 
the first right-hand term compensates 
``long tail'' of the second coming from (\ref{1}). 

This means that among eigen-values of PKO 
\,$\mck_\nu \hm =\mck_{1/t}$\,, in solutions of problem     
\begin{align}   \label{ev}  
\begin{array}{c}
  \mck_\nu \, R(V)  = -\,\lm\, R(V) \,\,\,,\, 
  \end{array}
\end{align} 
there are small eigen-values (e.v.) 
\begin{align}    \label{ev0}  
 \lesssim\,  \lm_*(\nu)\, = \, \nu\, \ln\, \frac 1{\tau_0\nu} \, =\, 
  \frac 1t\, \ln\, \frac t{\tau_0} \,\,  \,    
\end{align} 
with some \,$\tau_0$\,
(so that \,$\exp{[-\lm_*(\nu)\,t]}\propto 1/t$\,). 

Such e.v.s decreasing as QDF's width  
just are introducing to e.'s motion statistics 
``flicker'' (unrestrictedly low-frequency) fluctuations. 
 Next consider how it works and then discuss mathematical 
 origin of such e.v.

 {\bf 8}.\,  Being interested namely in flicker noise 
 it is reasonable to neglect DB violation 
 and thus difference between \,$\rho_{1/t}(V)$\, and  \,$\rho_0(V)$\,. 
 For that we may make symmetrization      
\begin{align}  \nonumber  
\begin{array}{c}    
w_\nu(V|V\pr) \,\rho_0(V\pr)   \Rightarrow \frac 12 
[\,w_\nu(V|V\pr)\,\rho_0(V\pr) \,  + \, \\  +\, 
\rho_0(-V)\, w_\nu(-V\pr |-V) ]  \Rightarrow  w_\nu(V|V\pr) \, \rho_0(V\pr)  \,    
  \end{array}
\end{align} 
thus 
subjecting PKO to operator equality
\begin{align}   \label{db}  
\begin{array}{c} 
  \mck_\nu \,\rho_0(V)  =   \rho_0(-V) \, \mck_\nu^ \dag  \,\, , \, 
\end{array}   
\end{align} 
where \,$\dag$\, is symbol of conjugation - 
transposition in the Sturm-Liouville sense plus 
inversion of velocity sign (so that operator  
\,$\rho^{-1/2}_0 \mck_\nu\rho^{1/2}_0$\, is self-conjugated). 
This property guarantees time-local DB observance \cite{vk,bk1,pufn}.  
Now  \,$\mck_\nu \rho_0(V) = 0$\, although QDF keeps its place.

{\bf 9}.\, Further recall that EP has argument \,$\gx$\, and    
\begin{align} 
\int_V \exp{ \{ \,t\,[- V \gx +  \mck_\nu ] \} } \,\rho_0(V) \, = \label{me} \,  
    \\ \ =\,   \nonumber      
  1 +\frac {M_2(t,\nu)}{2!}\,\gx^2 + 
 \frac {M_4(t,\nu)}{4!}\,\gx^4 + \,\dots\, \,  ,\,
\end{align}
where \,$M_n(t,\nu) $\, are statistical moments of displacement 
(path) of e. during time \,$t$\, under governing by given kinetic operator
of velocity \cite{oct}. It is taken into account that e.'s walk 
is spherically symmetric since gas of phonons is isotropic. 
Therefore eigen-functions (e.f.), or eigen-states (e.s.), 
\,$R(V)$\, in (\ref{ev}) can be sorted to odd and even. 
Let they be enumerated by indices ``1'' and  ``'2' respectively 
except equilibrium indexed with ``0''. 
 
Clearly, operator \,$V$\, as perturbation of \,$\mck_\nu$\, in (\ref{me})   
induces transitions only between its e.s.s with opposite 
parities. Hence in the second order of expansion over \,$\gx$\, 
on the left in (\ref{me}) we can write, with use of standard notations 
of perturbation theory (but leaving velocity's vector indices 
``in mind''), 
\begin{align} 
\frac {M_2}{2!} = \sum_1 \,V_{01}\, \frac
{t-(1- e^{-\lm_1 t})/\lm_1} {\lm_1}  
\,V_{10} \, \rightarrow Dt \,\, , \,   \label{m2} 
\end{align}
where \,$D$\, is diffusion coefficient (diffusivity) of e. 

We supposed that sets of matrix elements  
\,$V_{01},\, V_{10}$\, and e.v.s \,$\lm_1$\, are such that \,$D$\, 
is finite, i.e. e.'s diffusion is ``simple'' (not ``anomalous'').


{\bf 10}.\, Then at correspondingly large \,$t$\, 
the fourth order of the expansion tends to form    
\begin{align}  \label{m4} 
\frac {M_4}{4!} \equiv \frac {3 M_2^2 + C_4}{4!} 
\, \rightarrow \, \frac {D^2 t^2}2  +   
 \sum_{1,2,1^\prime } \, \frac {V_{01}V_{12}} {\lm_1}  \, 
  \,\times \\   \nonumber  \times\,\, 
\frac {t-(1- e^{-\lm_2 t})/\lm_2} {\lm_2} \,  
\frac {V_{21^\prime }V_{1^\prime  0}} {\lm_{1^\prime }}  \,\, . \,  
\end{align}
The triple sum here represents fourth-order cumulant 
of the e.'s path, \,$C_4 =M_4 -3 M_2^2$\,, characterizing, 
in comparison to \,$M_2^2$\,, degree of non-Gaussianity of 
of statistics of the walk \cite{ufn,1f,99,lufn,pufn}. 
 
If in (\ref{m4}) sets of even e.s.s of operator \,$\mck_\nu$\, 
and their e.v.s \,$\lm_2$\, were fixed, as for KO (\ref{uko}), 
then we would expected similar to (\ref{m2}) asymptotic 
 \,$C_4\propto t$\, giving  \,$C_4/ M_2^2\propto 1/t$\, 
 and asymptotically Gaussian walk. 
But our set   \,$\lm_2$\, is varying with observation time 
together with parameter  \,$\nu=1/t$\,, 
moreover, it is essentially varying as far as 
includes small \,$\nu$\,-dependent e.v.s (\ref{ev0}) 
(obviously belonging to even e.s.s in view of even parity  
of functions (\ref{kor0})-(\ref{1})). 
Because of them 
the fourth-order cumulant grows at large \,$t$\, not slower than  
\begin{align} 
C_4 = C_4(t,\nu=1/t)\, \propto\, \frac  {D^2\,t}{\lm_*(1/t)} 
\approx  \frac {D^2t^2}{\ln{(t/\tau_0)}} \,\,, \, \label{c4} 
\end{align}
while ratio \,$C_4/M_2^2$\, decreases not faster than logarithmically,
and the walk of e. is non-Gaussian even asymptotically.
It is convenient to describe such statistics in the language 
of diffusivity fluctuations with effective correlation function 
 \,$K_D(t) \hm \approx \pa_t^2 C_4/4!$\,  
\cite{ufn,157,1f,99,hbm,tbm,oct}, that is in our case   
\,$K_D(t) \hm \propto  D^2/\ln{(t/\tau_0)}$\, at \,$t\gg\tau_0$\,  
if not slower. 
 
Notice that inversely logarithmic tail of \,$K_D(t)$\,,
in company with realistic estimate of its weight, naturally appears
in statistical phenomenology of e. Brownian motion  
suggested in \cite{157,1f} (see also \cite{ufn,99,lufn}). 
Though generally (\ref{m4}) may produce also other dependencies 
more close to    
\begin{align}  \label{c4l} 
C_4\, \propto \,  D^2\,t^2   \, \,,\,
\end{align}
giving ``quasi-static'' fluctuations of diffusivity.  


{\bf 11}.\, Let us turn to the question 
how PKO \,$\mck_\nu$\,'s small e.v.s of order of \,$\nu$\,
can be explained from pure mathematical point of view. 
The answer comes with help of the perturbation theory (PT) of linear operators  
\cite{kato} if we treat PKO \,$\mck_\nu$\, as result of perturbation of  
KO \,$\mck_0$\,  
and 
pay attention to those principal circumstance that all e.v.s 
of operator \,$\mck_0$\, are multiply degenerated. 
Indeed, DDF in  (\ref{tp}) in place of QDF 
allows only transitions with precise conservation 
of sum \,$H_e +H_{ph}$\, of own energies of e. and PMs.
Therefore one and the same e.v. \,$\lm$\, in (\ref{ev}) with \,$\nu=0$\, 
can be associated with different e.f.s corresponding to different values of 
the ``integral of motion'' \,$H_e + H_{ph}$\, 
and specified, for instance, by their dependence on \,$E\hm = mV^2/2$\, 
on interval \,$0\hm\leq E\hm < E_0$\, where \,$E_0\hm\equiv \max\,\h\om_k$\, 
is maximal phonon energy.
Visual illustration is presented by special case of 
``ideally optical'' phonons with  
\,$\om_k \hm  =\om_0 \hm  \equiv\,$const\,.  
There  \,$E_0=\h\om_0\,$ and \,$\mck_0$\, connects only  
equidistantly by \,$E_0$\, separated points of \,$E$\, axis. 
Hence if  \,$R(V)$\, is a solution of (\ref{ev}) with \,$\nu=0$\, 
then \,$\Pi(E/E_0)\, R(V)$\,, where \,$\Pi(x)=\Pi(x\pm 1)$\, is 
arbitrary periodic function, 
also is solution of the same equation with same e.v., so that 
the degeneration is at least  countably-multiple. 

Perturbation \,$\mck_\nu-\mck_0$\, owing to QDF
introduces ``non-quantized'' transitions from any point 
of e.'s energy axis to all other its points and thus 
transitions between different \,$H_e +H_{ph}$\, values.
Physically it approaches the e.'s motion picture to reality 
where not  \,$H_e +H_{ph} $\, but the full energy 
\,$H_e +H_{ph} + H_{int}$\, is conserved. 
Formally according to PT it must lead to splitting 
of each e.v. to many different values \,$\propto\nu$\,. 
As the result, even a discrete e.v.s spectrum of parent operator 
transforms into arbitrarily dense one in whole permissible region, 
\,$\,0 \hm \leq \lm \hm < \lm_{max}\hm \approx  
\int_V\int_{V\pr} w_0(V\pr|V)\,\rho_0(V)$\, in our case. 
At that the zero e.v. connected to equilibrium serves as source  
of most interesting for us e.v.s \,$\sim \nu$\,.

{\bf 12}.\, We can ascertain the said by the example of optical phonons. 
For complete description of perturbed spectrum at lowest order of PT 
it would be necessary, for some full set of linearly independent 
functions \,$\Pi(x)$\,, to compose and solve corresponding 
secular equation. But for estimates of character and value of perturbation it is sufficient 
to apply simple formula of first-order PT for shift 
 \,$\vd\lm \hm =\lm \hm - \lm_0 $\, of e.v. \,$\lm_0$\, 
 as if it was non-degenerated: 
\begin{eqnarray}  \label{evs0}  
\vd\lm = -  \frac {\int_V \Pi^*\,F_0\, 
[\mck_\nu -\mck_0]\, \Pi\,F_0\,\rho_0} 
{\int_V  |\Pi|^2\, F_0^2\,\rho_0} \,\, , \,
\end{eqnarray}
where \,$\Pi=\Pi(E/E_0)$\,, and   
\,$F_0(V)\,\rho_0(V) =R_0(V)$\, is most smooth of  all e.f.s of     
\,$\mck_0$\, connected to \,$\lm_0$\, 
(in view of (\ref{db}) \,$\mck_0^\dag F_0 \hm =-\lm_0 F_0$\,).  
Choosing here 
\,$\Pi(x)\hm = \exp{(2\pi i n x)}$\, 
(with integer\,$n$\,) and treating QDF as sharp energy function  
as compared to \,$R_0(V)$\,, one can obtain  
\begin{eqnarray}    
\vd\lm \approx (\lm_{max}-\lm_0)   \nonumber  
[ 1 - \int_{-\infty}^\infty \delta_{\hbar\nu}(\ep)\, 
\cos{\frac {n\tau_0 \ep}{\h}} \,d\ep ] =\, 
 \\ \, \label{evs}  =\, 
(\lm_{max}-\lm_0)\, \min{(\tau_0 \nu |n|,1)} \,\, , \,\,\,\,\,
\end{eqnarray}
where 
\,$\tau_0  \equiv 2\pi\h/E_0$\,.     
Taking alternatively  
\begin{align} \nonumber 
\begin{array}{c}
\Pi(x) =1 \,\,\, {\tt if} \,\,\, 0< x \,{\tt mod}\,1\, \leq x_0 \,\,, \,\, 
\\ 
\Pi(x) =0 \,\,\, {\tt if} \,\,\, x_0< x \,{\tt mod}\,1\,\leq 1 \,\,   
\end{array}
\end{align}
with \,$x_0\leq 1 $\,, assuming for simplicity that \,$\h\nu/E_0 \ll x_0$\, 
and \,$\lm_0$\, lies at beginning of spectrum, 
and rejecting insignificant details, one finds  
\begin{eqnarray}    \label{lm} 
\vd\lm \approx  
2 \lm_{max} \,(1-x_0) \int_0^{x_0 E_0} \ep\, 
\delta_{\hbar\nu}(\ep)\, 
\frac {d\ep}{x_0 E_0} \, \approx \, \\ \, \nonumber \approx\, 
\lm_{max}\tau_0  \,\nu\, \frac {1-x_0} {\pi^2 x_0}\, 
\ln\, \frac {2\pi x_0}{\tau_0 \nu} \,\, . \, 
\end{eqnarray}

These test calculations quite highlight tendency of the  perturbation  
to expand \,$\lm_0$\, into ascending from \,$\lm_0$\, sequence
of e.v.s with step  
\,$\sim \lm_{max}\tau_0 \,\nu\, \ln\,({\tt const}/\tau_0 \nu) $\, 
(non-analiticity of expressions (\ref{evs})-(\ref{lm}) in respect to  
variable \,$\nu$\, is caused by that \,$\nu=0$\, is peculiar point 
of \,$\mck_\nu$\,). 
For us of most importance is perturbation of equilibrium state  
where  \,$F_0(V)\hm =$\,const\,, \,$\lm_0=0$\, and \,$\lm \hm =\vd\lm$\,.
Formulae (\ref{evs})-(\ref{lm}) show detachment of many small e.v.s from it 
including e.v.s \,$\approx \lm_*(\nu)$\, with  \,$\lm_*(\nu)$\, from (\ref{ev0}).

{\bf 13}.\, Consequently, among characteristic relaxation times 
of correlators and distributions of e.'s velocity there appear 
unbounded large ones \,$\sim \nu^{-1}=t$\,. 
Of course, in fact they only enter in definite types of correlators,   
since the perturbation does not destroy statistical isotropy of phonon field, 
and all e.s.s related to the equilibrium's degeneration and its removing 
are spherically symmetric along with \,$\rho_0(V)$\, 
(and with (\ref{kor0}) and (\ref{1})). 
Therefore they do not contribute to  (\ref{m2}) or to sums 
over \,$1$\, and  \,$1\pr$\, in (\ref{m4}) but contribute to sum over \,$2$\, 
in (\ref{m4}), moreover, this contribution becomes dominating with time. 
In the light of our estimates it is clear that the result is 
in between (\ref{c4}) and (\ref{c4l}). 
 
However, determination of numeric details in asymptotic like 
 (\ref{c4}) or (\ref{c4l}) requires more investigation of properties 
 of \,$\mck_0$\, and similar operators (not familiar, to the best 
 of our knowledge, even to mathematicians). In the past the physical kinetics 
had no need in it but novadays it becomes practically significant problem.

{\bf 14}.\, To resume, we suggested relatively simple approximation, 
(\ref{ap})-(\ref{ko}), of quantum dynamics of electron in phonon surroundings 
in terms of pseudo-kinetic operator (PKO) (\ref{ko})-(\ref{tp}) 
which differs from operators of convenient kinetics by evident 
taking account of finiteness of duration of real experiments. 
The PKO neglects many-phonon collisions but instead visually reveals 
uncertainty of electron's rate (coefficient) of diffusion (mobility) 
in the form of its flicker fluctuations (1/f-noise).

Common sense prompts \cite{ufn,157,1f,lufn} that related 
unbounded from above correlation times in essence signify absence 
of correlations in these fluctuations.  
The seeming paradox is resolved, - as was underlined in \cite{a2,195}
(see also \cite{99,pufn,lufn}), - at the expense of unusual statistics 
with Cauchy type probability distributions. 
 
Curiously, in the framework of PKO approximation the Cauchy distribution (CD) 
appears already in PKO itself in the form of quasi-delta-function 
(on the right in (\ref{kdf})). 
Due to it the PKO acts as if phonon energy undergoes random deviations 
from dispersion law obeying CD with width  \,$\h\nu \hm =\h/t$\,. 
At that those CD's peculiarity caused by its inversely quadratic tail  
is important that under its convolutions (reproducing CD) summation of 
widths of distributions takes place, 
i.e. independent (in the sense of probability theory) random quantities, 
each subject to CD, are summed like fully commonly dependent ones, 
which means their full indifference to averaging. 
Hence in the course of many collisions the random deviations of phonon energies 
do not suppress one another but produce effective deviation with 
amplitude (CD' width)  \,$\approx n(t)\,\h/t \hm \approx \h\lm_{max}$\, 
which does not decrease with time (\,$n(t)\hm \approx \lm_{max} t$\, 
is typical number of collisions during time \,$t$\,). 
This view onto PKO can be useful for numeric estimates and models 
of 1/f-noise, including generalizations of the PKO approximation 
to other systems.

\,\, 

\,\,

\vfill\eject


\begin{thebibliography}{99}

\bibitem{ufn}
 G.N. Bochkov, Yu.E.  Kuzovlev,    Sov. Phys. - Uspekhi\, {\bf  26} 829 (1983)   

\bibitem{157}
%
 
 Kuzovlev Yu.E. and Bochkov G.N.\, {\sl On origin and statistical characteristics of 1/f-noise.} \,  Preprint NIRFI No.\,157.\, 
 Gorkii (Nijniy Novgorod), USSR (Russia),  1982;\, arXiv:1211.4167 

\bibitem{1f}
%
 Kuzovlev Yu.E. and Bochkov G.N.,\, Radiophysics and Quantum Electronics\, {\bf 26} 228 (1983) 

\bibitem{a2}
%
 G.N.Bochkov and Yu.E.Kuzovlev,\, Radiophysics and Quantum Electronics\, {\bf 27} (9) 811 (1984) 

\bibitem{195}
 
 Bochkov G.N. and Kuzovlev Yu.E.\, {\sl Towards theory of 1/f-noise.} \,  Preprint NIRFI No.\,195.\, 
 Gorkii (Nijniy Novgorod), USSR (Russia),  1985 


\bibitem{i1}
 Yu.\,E. Kuzovlev,\,  Sov. Phys. - JETP\,  {\bf 67} (12) 2469 (1988); 
 arXiv:0907.3475

 \bibitem{99}  Yu.\,E.\,Kuzovlev,\, arXiv: cond-mat/9903350  

 \bibitem{tmf}
 Yu.\,E. Kuzovlev,\,  Theor. Math. Phys.\,  {\bf 160} 1301 (2009) 

\bibitem{hbm}  Yu.\,E.\,Kuzovlev,\, arXiv:1302.0373  
\bibitem{tbm}  Yu.\,E.\,Kuzovlev,\, arXiv:1207.0058   
 
\bibitem{lufn}
  Yu.\,E.  Kuzovlev,   Physics - Uspekhi\,  {\bf  58} (7)  719  (2015) 
 
\bibitem{gs}
  Yu.E.\,Kuzovlev,\, JETP Letters\, {\bf 103} (4) 234  (2016)

\bibitem{vk}
 N.G. van Kampen.\, {\sl Stochastic processes in physics and chemistry.}\, North-Holland Publ., 1992. 


\bibitem{july}  Yu.\,E.\,Kuzovlev,\, arXiv:1107.3240 
\bibitem{oct}  Yu.\,E.\,Kuzovlev,\, arXiv:1110.2502 

\bibitem{lp} 
 E.M. Lifshitz, L.P. Pitaevskii.\,  {\sl Physical kinetics}.\,  Oxford, Pergamon Press, 1981 


\bibitem{bk1} 
%
 G.N.Bochkov and Yu.E.Kuzovlev,\, Sov.Phys.-JETP {\bf 45}\, 125 (1977) 

\bibitem{pufn}
 G.\,N. Bochkov, Yu.\,E.  Kuzovlev,   Physics - Uspekhi  {\bf  56} (6)  590  (2013)  

\bibitem{kato}
%
 T. Kato.\,  {\sl Perturbation theory of linear operators.} \, Springer-Verlag, 1966 

 
\end{thebibliography}
\end{document}